\documentclass[]{aa}
\usepackage{times}
\usepackage{natbib}
\usepackage[dvips]{graphicx}
\usepackage{psfig,longtable,lscape}
\usepackage{psfig}
\usepackage{rotating}
\bibpunct{(}{)}{;}{a}{}{,}

\def\Swift{{\em Swift}}

\def\ltsima{$\; \buildrel < \over \sim \;$}
\def\simlt{\lower.5ex\hbox{\ltsima}}
\def\gtsima{$\; \buildrel > \over \sim \;$}
\def\simgt{\lower.5ex\hbox{\gtsima}}

\begin{document}

\title{Detectability of gamma-ray emission from classical novae with \Swift/BAT}
\author{F. Senziani\inst{1,2,3,4} \and G.K. Skinner\inst{5,6} \and P. Jean\inst{4} \and M. Hernanz\inst{7}}

\institute{Universit\`a di Pavia, Dipartimento di Fisica Nucleare e Teorica, Via Agostino Bassi 6, I--27100 Pavia, Italy
     \and Istituto Universitario di Studi Superiori (IUSS), V.le Lungo Ticino Sforza 56, I--27100 Pavia, Italy
         \and INAF-Istituto di Astrofisica Spaziale e Fisica Cosmica, Via Bassini 15, I--20133 Milano, Italy
     \and CESR/Universit\'e Toulouse III, 9, avenue du Colonel Roche, Toulouse 31028, France
     \and NASA/Goddard Space Flight Center, Code 661, Greenbelt, MD 20771, USA
     \and CRESST/Dept. of Astronomy, University of Maryland CP, College Park, MD 20742, USA
     \and IEEC-CSIC, Campus UAB, Fac. Cienc., C5-par, 2on, 08193 Bellaterra, Spain
}
\offprints{Fabio Senziani, senziani@iasf-milano.inaf.it}

\date{Received / Accepted}

\authorrunning{F. Senziani et al.}

\titlerunning{Detectability of gamma rays from novae with \Swift/BAT}

\abstract {Classical novae are expected to emit gamma rays during their explosions. The most important contribution to the early gamma-ray emission comes from the annihilation with electrons of the positrons generated by the decay of $^{13}N$ and $^{18}F$.
The photons are expected to be down-scattered to a few tens of keV, and  the emission is predicted to occur some days before the visual discovery and to last $\sim$2 days.
Despite a number of attempts, no positive detections of such emission have been made, due to lack of sensitivity and of sky coverage. } 
{Because of its huge field of view, good sensitivity, and well-adapted (14-200 keV) energy band, \Swift/BAT offers a new opportunity for such searches. 
BAT data can be retrospectively used to search for prompt gamma-ray emission from the direction of novae after their optical discovery.}
{We have estimated the expected success rate for the detection with BAT of gamma rays from classical novae  using a Monte Carlo approach.  Searches were  performed for  emission from novae occurring since the launch of \Swift.}
{Using the actual observing programme  during the first 2.3 years of BAT operations as an example, and sensitivity achieved,  we estimate the expected rate of detection of classical novae with BAT as $\sim 0.2-0.5 yr^{-1}$, implying that several should be seen within a 10 yr mission. The search for emission in the directions of the 24 classical novae discovered since the \Swift\ launch yielded  no positive results, but none of these was known to be close enough for this to be a surprise.  Detections of a recurrent nova (RS Oph) and a nearby  dwarf nova (V455 And) demonstrate the efficacy of the technique. }
{The absence of detections 
is consistent with the expectations from the Monte Carlo simulations, but the long-term prospects are encouraging given an anticipated \Swift\ operating lifetime of $\sim$10 years. }

\keywords {Gamma rays : observations - Nuclear reactions, nucleosynthesis, abundances - Stars: novae, cataclysmic variables - Stars: white dwarfs}

\maketitle

\section{Introduction}

\label{intro}
Classical novae are thought to be cataclysmic variables in which a white dwarf (WD) accretes material from a companion star until the conditions for  hydrogen ignition are reached on its surface  and a thermonuclear runaway sets in, leading to an explosion \citep{Starrfield1978}. As first noted by \citet{Clayton1974}, novae are expected to emit  gamma rays associated with the decay of the radioactive isotopes produced during the explosion. 
It is largely the decay of the short-lived $\beta^{+}$-unstable nuclei, such as $^{13}$N, $^{14,15}$O and $^{17,18}$F, that leads to a sudden
release of a great amount of energy, and the consequent envelope expansion, huge luminosity increase and mass ejection. 
The positrons emitted in the decay annihilate with electrons, producing a characteristic 511 keV line.
Comptonization of most of the 511 keV photons also leads to a continuum down to a few tens of keV. 
Among the short-lived isotopes, $^{13}$N and $^{18}$F (with lifetime $\tau$=862s and 158min respectively) are expected to be the most important contributors to the prompt gamma-ray emission since they decay when the envelope is starting to become transparent.
Later gamma-ray emission comes  from the decay of medium- and long-lived isotopes, such as $^{7}$Be (lifetime $\tau$=77days), $^{22}$Na ($\tau$=3.75yrs), and $^{26}$Al ($\tau$=10$^{6}$yrs). Photons are emitted by these nuclei with characteristic energies of 478 keV for $^{7}$Be, 1275 keV for $^{22}$Na and 1809 keV for $^{26}$Al. Because of their longer decay times, the decay rates of these isotopes is much lower than that of  $^{13}$N or $^{18}$F and they are not considered here. 

The gamma-ray emission depends critically on the nature and characteristics of the thermonuclear explosion, so its observation would provide a sensitive test of our theories of classical novae. However, to date no detection has been possible.
Because novae are unpredictable, detecting prompt gamma rays from novae requires a wide field of view (FOV) telescope that can continuously monitor the sky with good sensitivity over the appropriate energy range.
 Unsuccessful attempts have been made using BATSE on board CGRO \citep{Hernanz2000}, TGRS on board WIND \citep{Harris99,Harris2000}, and RHESSI \citep{Smith2004}. 
\citet{Hernanz_jose_2004} presented prospects for the detectability of the electron-positron annihilation continuum with IBIS/ISGRI telescope on board INTEGRAL. Despite the good sensitivity of the instrument, they found that a detection is expected only if a nova closer than 4-5 kpc lies by chance in the IBIS FOV within about 10 hours of the outburst temperature peak. 
The relatively small FOV (29$^\circ \times$ 29$^\circ$) of the IBIS telescope makes the probability of  this  very low and
indeed, it has not occurred during the five years since the launch of INTEGRAL.

The Burst Alert Telescope (BAT) on board the \Swift\ satellite \citep{Gehrels} offers a new opportunity to search for the prompt gamma rays from novae. 
The BAT satisfies all the requirements described above: it has a FOV of $\sim$ 2 sr (about an order of magnitude larger than that of IBIS) and covers $\sim$50\% of the sky each day. It has a sensitivity similar to that of IBIS/ISGRI and works  in an energy range (14-200 keV) that is well matched to  that of the expected nova emission. 

In this paper we investigate the detectability of the prompt gamma-ray emission from classical novae with the BAT.
We first discuss the expected fluxes in the BAT energy range based on the latest available nova models for the production of  $^{13}$N and $^{18}$F (\S  \ref{Jean_work}).
A description of the BAT performance and of the approach we adopted to search for prompt gamma rays from novae is given in \S \ref{instrument}.
In \S \ref{simulations} we  use a Monte Carlo approach to investigate the probability of detecting  the expected emission based on modelling the distribution of novae in the Galaxy and on the characteristics of the \Swift\  observing sequence.
Results of a retrospective search in the BAT data archive for gamma rays from 24 novae discovered during the first 3 years of  \Swift\ operations are presented in \S \ref{results}. 


\section{The expected prompt gamma-ray emission from novae}\label{Jean_work}


The gamma-ray flux expected during the explosion of a typical classical
nova depends on the amount of the isotopes $^{13}$N and $^{18}$F produced, and on
the probability that the gamma rays resulting from their decay can
escape. This flux has been computed with a suite of two numerical codes:
the hydrodynamical code developed to simulate a nova explosion and its
nucleosynthesis, from the accretion up to the explosion stages \citep{Jose1998},
 and the Monte Carlo code dealing with the production and
transfer of gamma rays in the expanding nova envelope \citep{Gomez-Gomar_et_al_1998}. 
The results depend on the mass of the WD and its chemical
composition (CO or ONe), the initial luminosity, the degree of mixing
between the WD core and the accreted envelope and the mass accretion rate;
typical masses adopted in the simulations are 1.25 M$_{\odot}$ for ONe-type and
1.15 M$_{\odot}$ for CO-type novae, with degrees of mixing of 50\%, accretion rate
2$\times$10$^{-10}$ M$_{\odot}$yr$^{-1}$ and $L_{in} =$ 1$\times$10$^{-2}$ $L_{\odot}$ (see \citet{Hernanz_et_al_2002}). The
results also depend critically on nuclear cross-sections, some of which
are poorly known (see \citet{Hernanz_jose_2006}). 
Important reductions
in the $^{18}$F yields with respect to models used in \citet{Gomez-Gomar_et_al_1998} have been reported in \citet{Hernanz_et_al_1999} and \citet{Coc_et_al_2000};
the nova models from these papers have been used here, and the expected
reduction in the gamma-ray flux due to the new $^{18}$F yields obtained
with the recent \citet{Chafa2005} cross sections has been applied as a
correcting factor in the second peak of the light curves (see below).
The successive reductions in the $^{18}$F yield along the past 10 years is
as follows, if we adopt the yields with the rates of \citet{Chafa2005}
as a reference:  a reduction by a factor of $\sim$8 with respect to \citet{Coc_et_al_2000}, a factor of $\sim$50 with respect to \citet{Hernanz_et_al_1999}),
and a factor of $\sim$500 with respect to \citet{Gomez-Gomar_et_al_1998}.

The situation concerning the more recent determinations of nuclear cross-section 
affecting $^{18}$F synthesis is not completely settled yet, 
but at least the uncertainties are becoming much smaller.

\begin{figure}
\centering
      \resizebox{\hsize}{!}{\includegraphics[angle=-90]{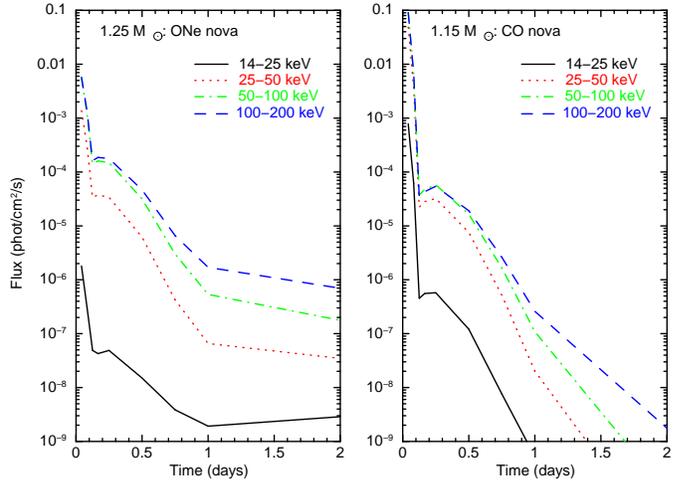}}
      \caption[]{\em Lightcurves in four energy bands for a 1.25 M$_{\odot}$ ONe-type nova (left) and for a 1.15 M$_{\odot}$ CO-type nova (right) based on the most recent available nova models and for a distance of 1 kpc.}
         \label{lc_Hernanz_ONe_CO}
\end{figure}

To assess the detectability with BAT of prompt gamma-ray emission from
novae, we ran the two codes mentioned for the above conditions of mass and
other parameters corresponding to a CO and an ONe nova.
The resulting predicted lightcurves in four energy bands are shown in Figure \ref{lc_Hernanz_ONe_CO}.
The strongest peak is due to the decay of $^{13}$N and reaches a maximum at $t \sim 1 hr$, the flux at earlier times being reduced by the envelope opacity. The second peak at $t \sim 6 hr$ is from the decay of $^{18}$F.  Depending on the nova speed class \citep{Gomez-Gomar_et_al_1998}, the peak gamma-ray emission is expected to occur some days before the visual maximum and hence probably  before the discovery.  Thus it is most likely to be detected if retrospective searches are made in the data from an instrument with good sky coverage.  The $^{18}$F emission is longer lasting  and so, with an instrument  making many different pointings, is easier to detect, provided the sensitivity is good enough. However  the  $^{13}$N peak is stronger and so detectable at greater distance.


\section{The \Swift/BAT instrument} \label{instrument}

As discussed in \S \ref{intro}, the BAT \citep{Barthelmy} on board \Swift\ is well adapted  to search for prompt gamma rays from novae.
It is a coded-mask telescope operating in the 
15-150 keV energy range.
The BAT detector plane is composed of 32768 pieces of CdZnTe (CZT), covering a 1.2 $\times$ 0.6 m sensitive area. 
A 2.7m$^{2}$  D-shaped coded mask,  with  $\sim 54,000$ lead tiles
arranged in a  random, 50\% open pattern, is mounted 1 metre above the detector plane.
The instrument FOV covers $\sim$ 1/6 of the sky in a single pointing. Within the field of view the sensitivity depends on the `coding fraction' which varies from 1 (`fully coded') in the central 0.5 sr of the field, to 0 at its limits. The  FOV with a coded fraction greater than 0.1 is about 2.2 sr (1.5 sr with coding $>$0.5).

For exposures of duration $\Delta t$, the 5$\sigma$ sensitivity (ph/cm$^2$/s) for a point source at a coded fraction of $n_{c.f.}$ is given by
\begin{equation}
S(E) = \frac{5 \cdot \sqrt{r_{B}(E)}}{(A_{det}/2) \cdot \eta(E) \cdot f_{m} \cdot \sqrt{n_{c.f.} \cdot \Delta t} }
\label{sensitivity_eqn}
\end{equation}
where $r_{B}(E)$ and $\eta(E)$ are the background count rate in the whole detector and the detector efficiency in the energy band under consideration.
$A_{det}$ is the detector plane area and $f_{m}$ (equal to 0.73 for BAT) is a factor 
which takes into account the  size of the detector pixels relative to the mask elements \citep{Skinner2008}.

The \Swift\ observing strategy has important implications for the search for prompt emission from novae.
As the \Swift\ satellite is specifically designed to catch and study  gamma-ray bursts (GRBs), the BAT field is generally directed away from the earth. Several different 
pointings are made in each spacecraft orbit (96 min). The directions are chosen such that its companion narrow-field instrument, the XRT (X-ray Telescope), can observe particular objects or follow-up GRB afterglows. As the XRT 
targets are well distributed over the sky, the BAT is able to collect  `survey mode' data from a large fraction of the sky every day. In survey mode, while awaiting GRBs, data are accumulated on board in arrays, 
called Detector Plane Histograms (DPHs), that are periodically  sent to the ground. 
A DPH contains a 80-channel spectrum for each of the 32768 pixels of the detector plane.
Such histograms are integrated over 450 seconds, or sometimes less if the integration is terminated for operational reasons such as the start of a spacecraft slew or the approach to the South
Atlantic Anomaly.
On board software searches  BAT event rates and images for increases corresponding to GRBs,
working over a range of time scales (see \citet{Gehrels} and \citet{Fenimore2003}).
The prompt gamma-ray emission from a nova  could in principle trigger the GRB detection system and lead to a spacecraft  repointing, but it would probably be  too faint and long-lasting to be detected in this way.
However, once it is known that a nova has occurred at a particular location, a more sensitive retrospective search is possible.  Thus the survey mode DPH data provides a valuable resource for nova emission searches.

A particular direction on the sky is typically within the field of view many times per day, sometimes in an irregular pattern, sometimes with consecutive observations spaced by the 96 min orbital period.
This  sampling pattern is particularly relevant for
the detection of novae because their peak flux could be missed during the gaps.


\section{Prospects for detection}\label{simulations}


\subsection{Monte Carlo simulations}\label{MC_sim}

Whether or not prompt gamma-ray emission from a particular nova can be detected with BAT depends in a complex way on the distance and gamma-ray luminosity of the nova and how the gamma-ray light curve is sampled by the sequence of pointings for which it lies within the field of view.   
To estimate the probable rate of detection of novae with  BAT, we generated a sample of imaginary novae with a Monte Carlo simulation, that were checked against a typical observing sequence. For the observing programme  we used the  actual BAT pointings made during the first 2.3 years of operation. 

\subsubsection{Nova distribution models}

Novae were simulated based on each of  three different models of 
spatial distribution in the Galaxy. The models were those used by \citet{Jean2000}, 
excluding their oldest population model. 
Details of the three adopted models are summarised in Table \ref{table_Jean2000}.
Each has a different distribution between the novae in the bulge and those in the disk. 

\begin{table*}[htbp]
\caption{\em Models of the nova spatial distribution in the Galaxy used for the Monte Carlo simulations \citep{Jean2000}. }
\label{table_Jean2000}
\begin{center}
\footnotesize{
\begin{tabular}{lll}
\hline \\
\multicolumn{3}{l}{{\bf Model 1} \citep{KDF91}:} \\
\multicolumn{3}{l}{$\rho_{h}$ = 3.0 kpc and $z_{h}$ = 0.170 kpc. $K_{0}$ is the modified Bessel function} \\
\hline \\
Disc $n(z,\rho)$ & $= n_{d} exp \Big( -\frac{|z|}{z_{h}} - \frac{\rho}{\rho_{h}} \Big)$   & \\[6pt]
Bulge $n(R)$     & $= n_{s} 1.04 \times 10^{6} \Big( \frac{R}{0.482} \Big)^{-1.85}$  &  $R \leq 0.938 kpc$ \\[6pt]
                 & $= n_{s} 3.53 K_{0} \Big( \frac{R}{0.667} \Big)$  &  $R \geq 0.938$ \\[6pt]
                 & $= 0$  &  $R \geq 5 kpc$ \\[6pt]
\hline \\
\multicolumn{3}{l}{{\bf Model 2} \citep{VdK90}:} \\
\multicolumn{3}{l}{$\rho_{h}$ = 5.0 kpc and $z_{h}$ = 0.30 kpc} \\
\hline \\
Disc $n(z,\rho)$ & $= n_{d} exp \Big( -\frac{|z|}{z_{h}} - \frac{\rho}{\rho_{h}} \Big)$   & \\[6pt]
Bulge $n(R)$     & $= n_{s} 1.25 \Big( \frac{R}{R_g} \Big)^{-6/8} exp \Big[ -10.093 \Big( \frac{R}{R_g} \Big)^{1/4} + 10.093 \Big]$  &  $R \leq R_g$ \\[6pt]
                 & $= n_{s} \Big( \frac{R}{R_g} \Big)^{-7/8} \Bigg[ 1-\frac{0.0867}{ \big( \frac{R}{R_g} \big)^{1/4}} \Bigg] exp \Big[ -10.093 \Big( \frac{R}{R_g} \Big)^{1/4} + 10.093 \Big]$  &  $R \geq R_g$ \\[6pt]
\hline \\
\multicolumn{3}{l}{{\bf Model 3} \citep{DJ94}:} \\
\multicolumn{3}{l}{$\rho_{h}$ = 5.0 kpc and $z_{h}$ = 0.35 kpc} \\
\hline \\
Disc $n(z,\rho)$ & $= n_{d} exp \Big( -\frac{|z|}{z_{h}} - \frac{\rho}{\rho_{h}} \Big)$   & \\[6pt]
Bulge $n(R)$     & $= \frac{n_{s}}{R^{3}+0.343}$ &  $R \leq 3 kpc$ \\[6pt]
                 & $= 0$  &  $R \geq 3 kpc$ \\[6pt]
\hline \\

\end{tabular}}
\end{center}
{\em $R$ is the distance from the Galactic Centre, $z$ is the distance perpendicular to the Galactic plane and $\rho$ is the galactocentric planar distance. The distance from the Galactic Centre to the Sun is $R_g = 8 kpc$, $n_{s}$ and $n_{d}$ are the normalisation factors for the bulge and the disc respectively (in kpc$^{-3}$). The proportions of novae in the bulge are 0.179 (Model 1), 0.105 (Model 2) and 0.111 (Model 3).}
\end{table*}

\subsubsection{The rate of novae}\label{rate_of_novae}

The rate of occurrence of classical novae in the Galaxy has been variously estimated as  $35 \pm 11$ yr$^{-1}$   \citep{Shafter_1997}, $41 \pm 20$ yr$^{-1}$   \citep{Hatano1997}  and $34^{+15}_{-12}$  yr$^{-1}$ \citep{ Darnley06}.

\begin{figure}[htbp]
      \resizebox{\hsize}{!}{\includegraphics{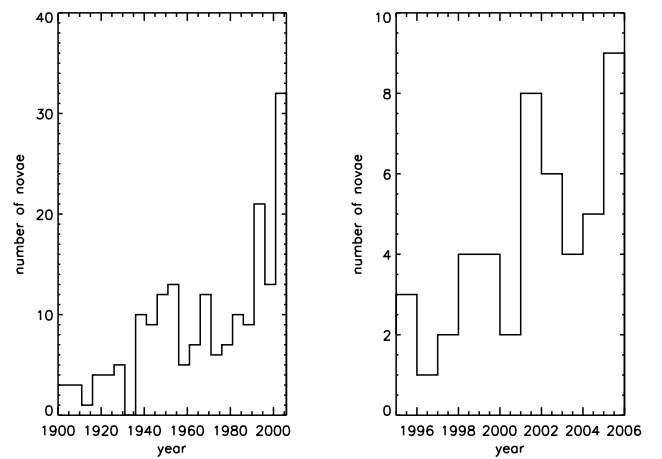}}
      \caption[]{\em Histogram of the novae discovered in 5 years intervals from 1901-2005 (left) and per year from 1995-2005 (right).}
         \label{rate_novae.ps}
\end{figure}

However, to be the subject of a retrospective search in BAT data, a nova has first to be discovered in the optical domain. Thus the 
relevant rate is strictly the frequency of {\sl reported}  classical novae.
Interestingly,  the discovery of a classical nova depends mainly 
on the work of amateurs astronomers. 
A very useful catalogue is available on the web, containing all novae discovered up to February 1, 2006 
(see \citet{Downes2001} and the web-site {\em http://archive.stsci.edu/prepds/cvcat/index.html}). 
From that, we selected all those objects classified as  classical novae. 
The number discovered in 5 year intervals over  the last century 
is shown in the left panel of Figure \ref{rate_novae.ps}. 

The general rising trend  in the number of novae discovered per year is certainly related to 
the steadily increasing activities of amateurs. 
The righthand panel of Figure \ref{rate_novae.ps} shows the number  discovered 
during each of the last ten years.
The rate has recently risen to 
($\sim$ 10 yr$^{-1}$), but is still far from the  
predicted Galactic nova rate.  This is mainly due to interstellar extinction, though at any one time a part of the sky is too close to the sun for a nova to be observed.

\subsubsection{White dwarf composition}

According to the calculations described in \S \ref{Jean_work}, at very early times CO novae release more
gamma rays related to $^{13}$N decay than ONe ones, so for the simulations we make an assumption about
the relative numbers of ONe and CO novae. We note however that this difference
is probably not a general trend, since it is strongly related to the
efficiency of convection in the outer nova layers, and thus a range of
fluxes between those presented here is expected, making the results less dependent on the assumption made.


The classification of a nova as type ONe or CO  relies on spectroscopic observations and is not always available.  \citet{Livio_Truran_1994} estimated that  ONe novae represent between 11\% and 33\% of galactic novae.   The results presented here are based on a mix of 25\% ONe and 75\% CO novae, both in the disc and in the bulge, but  results are also given for  ONe and for CO novae separately.

We note that the expected gamma-ray flux from a nova depends mainly on the WD mass 
(see \citet{Hernanz_et_al_2002}) and  the fluxes shown in Figure \ref{lc_Hernanz_ONe_CO} were calculated assuming single, typical,  WD masses  for ONe and for CO novae.
In this respect,  the approach adopted here is  a somewhat simplified one.

\subsection{Monte Carlo results}

\label{mc_results}
\begin{figure}[htbp]
      \resizebox{\hsize}{!}{\includegraphics{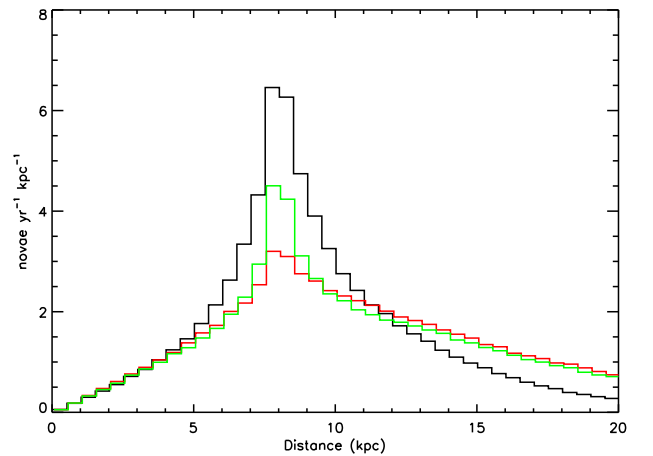}}
      \caption[]{\em Distribution of the distances from the Sun of the simulated novae. Black, red and green histograms correspond to  novae distributed according to models 1, 2 and 3 respectively (see Table \ref{table_Jean2000}).}
         \label{distrb_distanze_novae_simul_3modelli.ps}
\end{figure}

For each distribution model, the positions in the Galaxy and the time of explosion of  350000 imaginary novae were generated, finding for each one a distance from the Sun and the sky coordinates. All of the explosion times chosen were during the first 2.3 years of BAT observations, for which we know the actual observing sequence and observing efficiency.  Each nova was allocated to a class (ONe or CO). The distributions of the distances from the Sun of the simulated novae is shown in Figure \ref{distrb_distanze_novae_simul_3modelli.ps} for each of the three models.

We then assessed  whether BAT would have seen the  gamma rays from the novae.
If and when during the 48 hours following the nova it  would have been within the BAT field of view, the sensitivity for a source at the corresponding position in the FOV was calculated using Equation \ref{sensitivity_eqn}. 
A nova was assumed to have been detected if the expected flux from Figure \ref{lc_Hernanz_ONe_CO}, integrated over the observation time and  scaled  to the distance of the simulated nova, exceeded the 5$\sigma$ sensitivity.

\begin{figure}[htbp]
      \resizebox{\hsize}{!}{\includegraphics{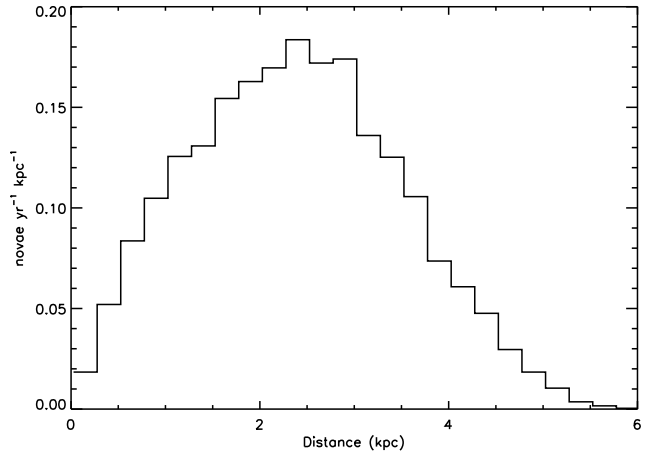}}
      \caption[]{\em Distribution of the distances from us of the novae detected in gamma rays in the Monte Carlo simulation. Model 1 was assumed.}
         \label{distrb_distanze_novae_detect_mod1.ps}
\end{figure}

Figure \ref{distrb_distanze_novae_detect_mod1.ps} shows the distribution of the distances of the detectable novae  according to the Monte Carlo simulation.  Although close novae are more easily detected, more distant ones are more numerous, leading to a peak at $\sim$2-3 kpc, although detection of novae at distances up to  $\sim$5 kpc is possible in favourable circumstances.

These results allow one to calculate the probability that, if a nova explodes somewhere in the Galaxy, the gamma rays would be detectable with the BAT. This can be multiplied by an appropriate nova rate (\S \ref{rate_of_novae}) to give the expected detection rate.  

The results for model 1 of Table \ref{table_Jean2000} are given in Table \ref{rates_detect} . We have not given explicitly the results for the other two models because they are essentially identical.
In calculating the detection probability, it is considered that a nova would be detected if it would have been seen at a significance level greater than 5$\sigma$ using any of the following integration timescales: single DPH ($\sim$ 13 minutes), 1 h, 3 h, 6 h, 12 h and 24 h. Usually it is the single DPH or 1 h timescales that prove most sensitive.

\begin{table}[htbp]
\caption{\em The probability that a novae is detectable with BAT, for each of  four energy bins. Results are for the spatial distribution of model 1 in Table \ref{table_Jean2000}. Other distribution models give very similar results  (see text for details).} 
\label{rates_detect}

\begin{center}
\scriptsize{
\begin{tabular}{ccccc}
\hline \\
Nova type & \multicolumn{4}{c}{Detection probabilities} \\
proportions & 14-25 keV & 25-50 keV & 50-100 keV & 100-200 keV \\
\hline \\
100\% ONe                    & 0        &   0.02 \%   &   0.12\% &   0.10\% \\
100\% CO		     & 0.007\%   &   0.82\%   &   1.95\%  &   1.59\% \\
25\% ONe + 75\% CO	     & 0.004\%	&   0.63\%   &   1.55\%  &   1.25\% \\
\hline \\
\end{tabular}}
\end{center}
\end{table}

Table \ref{rates_detect} shows that the 50--100 keV band is the most promising one and that 
  the probability of  detection of a particular nova is only about 1.5\%.  Nevertheless,  with a nova rate of 35  yr$^{-1}$ and a hoped-for mission life of 10 yr, this implies an expectation number of 5.2 detected novae.  However, as discussed in \S \ref{rate_of_novae}, a nova must be discovered optically in order to perform the retrospective search for gamma-ray emission. At first sight this would imply that an effective rate of $\sim10$ yr$^{-1}$ should be multiplied by 1.5\%, suggesting only 1.5 over a 10 yr \Swift\ mission.  
 
\begin{figure}[htbp]
      \resizebox{\hsize}{!}{\includegraphics{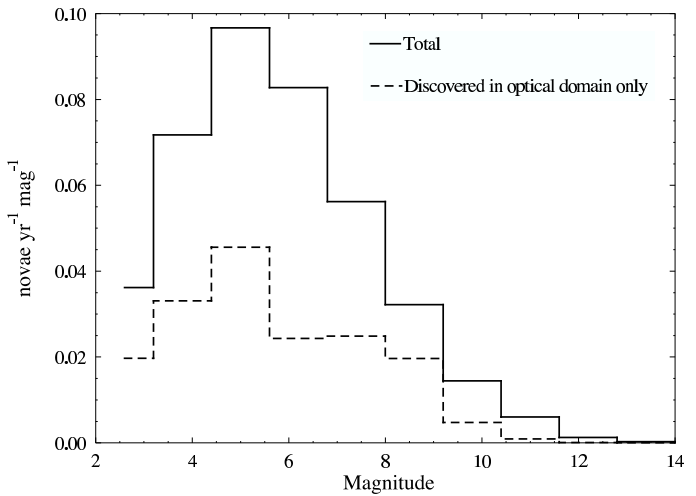}}
      \caption[]{\em Distribution of the magnitude of the novae detected in gamma rays in the Monte Carlo simulation (solid line). The dashed line represents the same distribution weighted for the probability of optical discovery of a nova (see text for details).}
         \label{distrb_magn_novae}
\end{figure}

However the novae that are discovered optically are preferentially those that are close, both because they are brighter and because interstellar absorption tends to be less. 
We have examined the distribution in $m_V$ of the discovered novae in \citet{Downes2001} catalog, considering novae discovered between 1985 and 2005. This can be compared with the corresponding distribution that would be expected based on our simulated nova population, taking the absolute visual magnitude of a typical nova as $-$8 and allowing for galactic extinction based on \cite{Hakkila1997}. We conclude
that although the above rates imply that only about 1 nova in 3$-$4 is observed,  the probability of optical discovery of a nova of apparent magnitude  $m_V <$ 9 is about  50\%, roughly independent of magnitude.   For $m_V >$ 9 it decreases rapidly and is essentially zero for  $m_V>14$.
 We used such a probability distribution to weight the distribution in magnitude of the novae detected with BAT among those of our simulated population.
Both the weighted and the unweighted distributions of the novae detected with BAT are plotted in Figure \ref{distrb_magn_novae}. 


We conclude that based on the assumptions made here 2$-$5 novae should be detectable in gamma rays during a 10 yr mission. 

\section{Observational results}\label{results}

\subsection{Classical novae since the launch of \Swift }\label{uplim}

The classical novae discovered during the first $\sim$3 years after the \Swift\ launch are listed in
Table \ref{17_novae}.  Of these,  7 novae have been identified as CO type. Distance measures are available in the literature  for only  5 of the novae. Where sufficient data are available we have attempted to make our own distance estimates using  the peak magnitude, $m_{V}$, the time to decay by 2 magnitudes,  $t_{2}$, the reddening $E(B-V)$ and the  MMRD relationship of \citet{della_Valle_Livio_1995}. These are given in  Table \ref{17_novae_final_table}. $m_{V}$,  and  $t_{2}$, were estimated from  AAVSO  lightcurves (see {\em http://www.aavso.org/}).  $E(B-V)$ values were taken from IAU Circulars when available, otherwise the extinction map of  \citet{Hakkila1997} was used (see Table \ref{17_novae_final_table}). The derived distances must be treated with some caution because of the large uncertainty in the values used  and in the MMRD relationship adopted.

\begin{table*}[htbp]
\caption[]{\em The 24 classical novae 
examined in the search for the prompt gamma-ray emission.}
\label{17_novae}

\begin{center}
\begin{tabular}{lllllccl}
\hline \\
Nova & RA & DEC & Date of discovery & Last prediscovery & Nova type & Distance from    & References* \\
     & (deg)   & (deg)    &   (UTC)   & (UTC)      &           & literature (kpc) &            \\
\hline \\
\object{V2361 Cyg} &  302.329  & 39.814 & 2005 Feb 10.850 & 2005 Feb 6 & CO,FeII  &              & 1,2,3,4,5,6,7 \\
\object{V382 Nor} &  244.936  & -51.581 & 2005 Mar 13.309 & ? & FeII     &              & 8,9 \\
\object{V378 Ser} &  267.3523 & -12.999 & 2005 Mar 18.345 & ? & CO,FeII  &              & 10,11,12,13,6 \\
\object{V5115 Sgr} &  274.246  & -25.944 & 2005 Mar 28.779 & 2005 Mar 27.464 &        &              & 14,15 \\
\object{V1663 Aql} &  286.302  & 5.236 & 2005 Jun 9.240 & ? & FeII      & 2.9 $\pm$ 0.4 & \citet{Boyd_2006} \\
          &              &          &                &                 &           & 5.5 $\pm$ 1  & \citet{Lane_et_al_2006} \\
          &              &          &                 &                &           &  7.3-11.3    &  \citet{Poggiani_2006}\\
          &              &          &                   &              &           &              & 16,17,18 \\
\object{V5116 Sgr} &  274.461  & -30.442 & 2005 Jul 4.049 & 2005 Jun 12 &        &              & 19,20,21,22 \\
\object{V1188 Sco} &  266.089  & -34.276 & 2005 Jul 25.284 & 2005 Jul 20 & FeII(?) &              & 23,24,25,26 \\
\object{V1047 Cen} &  200.207  & -62.630 & 2005 Sep 1.031 & 2005 Aug 12.050 &        &              & 22 \\
\object{V476 Sct}  &  278.020  & -6.726  & 2005 Sep 28.088 & 2005 Sep 24.629 & FeII & 4 $\pm$ 1 & \citet{Munari_et_al_2006} \\
          &           &        &                  &                &          &           & 27,28,29 \\
\object{V477 Sct} &  279.678  & -12.271 & 2005 Oct 11.026 & 2005 Oct 7.055 & He/N     & 11 $\pm$ 3.6 & \citet{Munari_2006} \\
         &              &            &                 &                 &          &              & 30,31 \\
\object{V2575 Oph} &  263.304  & -24.351 & 2006 Feb 8.379 & 1991 Aug 11 & CO,FeII &                   & 32,33,34,35 \\
\object{V5117 Sgr} &  269.719  & -36.793 & 2006 Feb 17.370 & 2006 Feb 5.36 & FeII    &              & 36,37 \\
\object{V2362 Cyg} &  317.885  & 44.801 & 2006 Apr 2.807 & 2006 Mar 28 & CO,FeII  & $\sim 1.5$    & \citet{Atel792} \\
          &              &            &                &                &          &     5-12      &  \citet{Atel795} \\
          &              &           &                 &                &          &               & 38,39,40,34,41,42,43 \\
\object{V2576 Oph} &  258.887  & -29.161 & 2006 Apr 6.565 & ? & CO,FeII &               & 44,34,45 \\
\object{V1065 Cen}  &   175.793  &  -58.067 &  2007 Jan 23.354  & 2007 Jan 15.36 &     &               &  46,47  \\
\object{V1280 Sco}  &   254.420  &  -32.343 &  2007 Feb 4.854   & 2007 Feb 2.866 & FeII  &             &   \citet{Swank2007ATel} \\
           &               &             &                   &                 &       &             &\citet{Osborne2007ATel}\\
           &               &             &                   &                 &       &             &  48,49,50,61 \\
\object{V1281 Sco}  &   254.247  &  -35.363 &  2007 Feb 19.859  & 2007 Feb 14.8575 &       &             & 51,52,62 \\
\object{V2467 Cyg}  &   307.052  &  41.810  &  2007 Mar 15.787  & 2007 Mar 12.796 & FeII &  1.5-4       & \citet{Steeghs2007ATel} \\
           &               &             &                   &                 &       &            &  53,63,64,65  \\ 
\object{V2615 Oph}  &   265.683  & -23.676  &  2007 Mar 19.812  & 2007 Mar 17.82  & CO,FeII &        &  54,62 \\
\object{V5558 Sgr}  &   272.577  &  -18.781 &  2007 Apr 14.777 & 2007 Apr 9.8  & FeII &  &  55,56,66,67\\
\object{V598 Pup}   &   106.428  &  -38.245 &  2007 Jun 5.968  & 2007 Jun 2.978 & CO  &  &  \citet{Read2007ATel} \\
                    &            &          &                  &               &       & &  \citet{Torres2007ATel} \\
                    &            &          &                  &               &       & &  \citet{Read2007ATel2} \\
                    &            &          &                  &               &       & &  76,77,78 \\
\object{V390 Nor}   &   248.048  &  -45.154 &  2007 Jun 15.086 & 2007 May 20.1 & FeII    &  &  57,68 \\ 
\object{V458 Vul}   &   298.601  &  20.880  &  2007 Aug 10.01  & 2007 Aug 4   &        &  &  69,70,58,71,72,73 \\
\object{V597 Pup}   &   124.075  &  -34.257 &  2007 Nov 14.23  & 2007 Nov 11.22 & FeII   &  &  59,60,74,75 \\
\hline \\
\end{tabular}

\end{center}
{\em * For each nova references to the first few IAU Circulars are given: (1) \#8483; (2) \#8484; (3) \#8487; (4) \#8511; (5) \#8524; (6) \#8529; (7) \#8641; (8) \#8497; (9) \#8498; (10) \#8505; (11) \#8506; (12) \#8509; (13) \#8527; (14) \#8502; (15) \#8523; (16) \#8540; (17) \#8544; (18) \#8640; (19) \#8559; (20) \#8561; (21) \#8579; (22) \#8596; (23) \#8574; (24) \#8575; (25) \#8576; (26) \#8581; (27) \#8607; (28) \#8612; (29) \#8638; (30) \#8644; (31) \#8617; (32) \#8671; (33) \#8676; (34) \#8710; (35) \#8728; (36) \#8673; (37) \#8706; (38) \#8697; (39) \#8698; (40) \#8702; (41) \#8731; (42) \#8785; (43) \#8788; (44) \#8700; (45) \#8730; (46) \#8800; (47) \#8801; (48) \#8803; (49) \#8807; (50) \#8809; (51) \#8810; (52) \#8812; (53) \#8821; (54) \#8824; (55) \#8832; (56) \#8854; (57) \#8850; (58) \#8863; (59) \#8895; (60) \#8896; (61) \#8845; (62) \#8846; (63) \#8848; (64) \#8888; (65) \#8905; (66) \#8874; (67) \#8884; (68) \#8851; (69) \#8861; (70) \#8862; (71) \#8878; (72) \#8883; (73) \#8904; (74) \#8902; (75) \#8911; (76) \#8898; (77) \#8899; (78) \#8901.}
\end{table*}

\subsection{Analysis}\label{analysis}

For each nova, we selected all the BAT observations having the position of the nova in the FOV  from 20 days before to 20 days after the nova discovery.
The relatively long time interval  takes into account the uncertainty in the time difference 
between the expected prompt gamma-ray emission and the nova discovery (see section \ref{Jean_work}) and also provides a baseline for comparison.

The DPH files (see \S \ref{instrument} for details) for the selected datasets were
analysed with a procedure combining public tools\footnote{Version 2.5 of the \Swift\ software (see http://swift.gsfc.nasa.gov/docs/software/lheasoft) was used.} 
and a special purpose fitting programme.
The data were filtered to exclude periods of high background rates, source occultation, etc.
Typically, a DPH file contains information for several consecutive pointings, each lasting 450 sec or less, 
with  an average  total exposure time of $\sim$13 min. As the peak flux is expected to occur on a timescale 
of $\sim$1 hr (see Figure \ref{lc_Hernanz_ONe_CO}), all of the pointings within a single DPH were combined.

The data were analysed in four energy bands, 14$-$25, 25$-$50, 50$-$100, and 100$-$200 keV, using a procedure that fits simultaneously a series of background models, the intensities of known strong sources and that of a source at the position of the nova.   

For each nova and energy bin, the  flux estimates were used to form a lightcurve.
All the fluxes measured with a signal to noise ratio (SNR) greater than 4 in one energy band 
or  having SNR greater than 2.5 in at least two energy bands were selected for further analysis. 
For each DPH in which such high points were found, an image of the FOV was generated using the   {\em batfftimage} public tool
and searched (with {\em batcelldetect}) for the evidence of the nova, setting a threshold of  5$\sigma$ level.
The requirement for detection of a signal in both the lightcurve and the image improves the robustness of any detection.

If no such double detection was found, the procedure was repeated  on lightcurves 
rebinned with different timebins (1h, 3h, 6h, 12h, 24h). To obtain images on the longer timescales and with different spacecraft attitudes, a special purpose programme was used to mosaic those from individual DPHs.


\subsection{Results of the search}

\begin{figure*}[htbp]
\centering
      \resizebox{\hsize}{!}{\includegraphics[angle=-90]{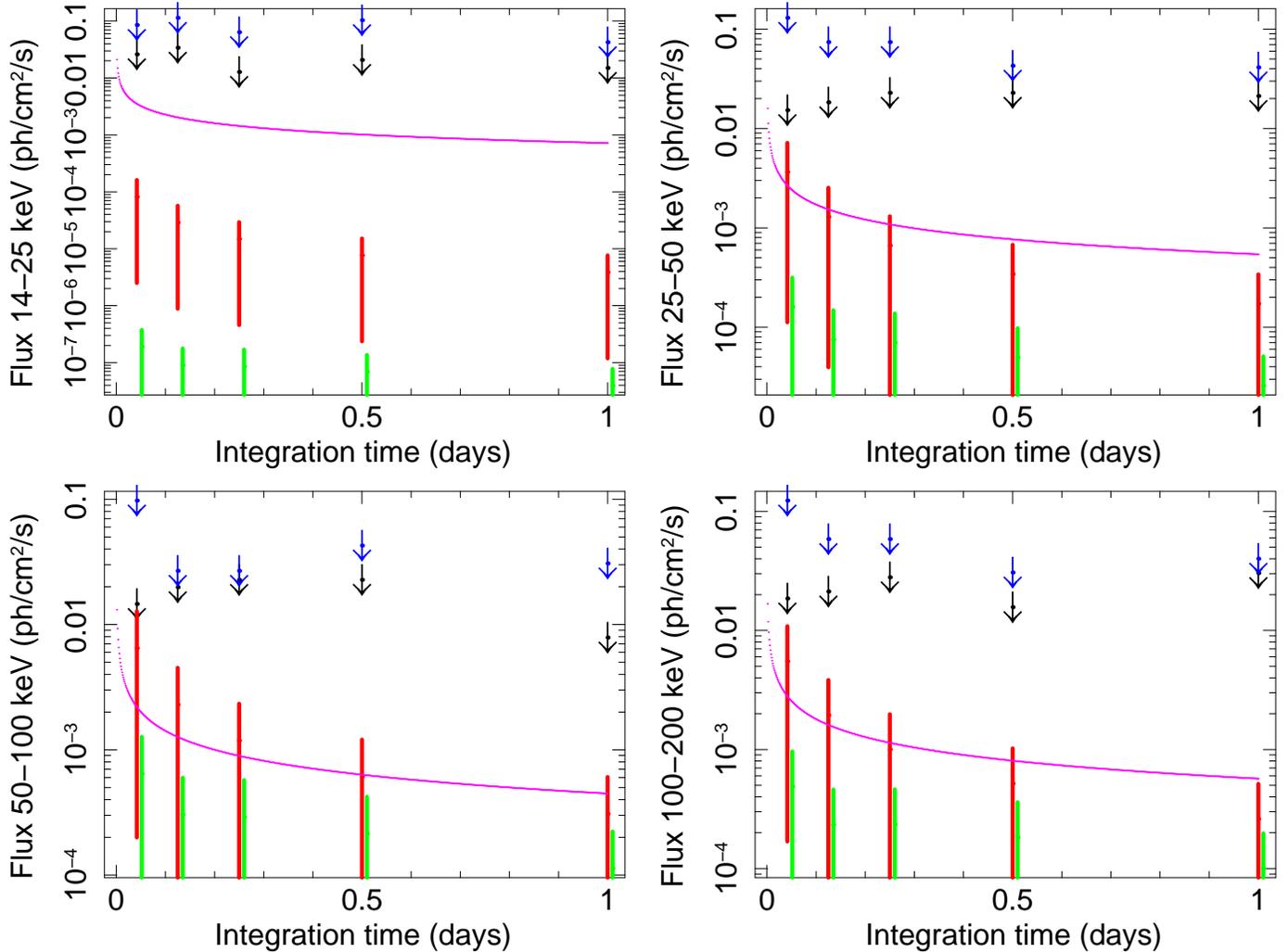}}
      \caption[]{\em Average expected flux as a function of integration time for the nova \object{V2362 Cyg}. Red  and green bars are supposing a CO type and a ONe type nova respectively.   
 Upper limits from BAT are indicated by  arrows, for percentage coverage of possible onset times  $P=$10\% (black) and  $P$=50\% (blue).       The  purple curve represents the ideal 5$\sigma$ BAT sensitivity if the source were observed continuously in the fully coded FOV. As expected, for long integrations the sensitivity is better, but the average flux lower.}
         \label{V2362_Cyg_FINALE_flux_vs_binsize.ps}
\end{figure*}

No evidence was found for prompt  gamma-ray emission from any of the 24 classical novae listed in Table \ref{17_novae}. 
No points satisfied both the above trigger conditions (lightcurve plus image detection).
The distribution of the complete ensemble of SNR measurements ( $\sim$750000 points, including all the single DPH light curves and all the energy bands)  agrees very well with a random
distribution with an average value of zero. The same is true for the rebinned light curves.
We therefore conclude that in each case either the expected gamma-ray emission from the novae occurred when the object was
not in the BAT FOV or the distance was too large for the predicted emission to be detected or the gamma-ray luminosity of the nova is  
weaker than expected.

To quantify the limiting distance below which a particular nova could be detected by BAT assuming the fluxes 
presented in section \ref{Jean_work}, we first estimated a limiting flux for each point of the lightcurves for each nova. Lightcurves at different energies and timescales were considered separately
in order to identify which combination of energy bin and timescale was most constraining.

Not only is the BAT coverage not continuous, but it also varies both in effective exposure 
and in the observing sequence from a source to another.
As pointed out above, the peak of the expected  gamma-ray flux of a nova could be missed during a coverage gap. Because the exact time of the peak is unknown, it is only possible to give a probability that  emission at a particular level would have been detected, assuming that the peak occurred at a random, unknown, time. What can be done is to say that for a certain percentage, $P$,  of  possible onset times, then the sensitivity and the coverage are such that the nova would have been detected had it been emitting at the level discussed in \S \ref{Jean_work} and had it been closer than a distance $d$. Where a distance estimate is available it can then be compared with $d$.
The calculation was done for $P$=10\% and for $P$=50\%.

We illustrate this process in the case of  nova \object{V2362 Cyg} in Figure \ref{V2362_Cyg_FINALE_flux_vs_binsize.ps}.
BAT upper limits are indicated with an arrow. 
Black symbols show the limits which can be placed on the flux for a percentage $P$= 10\% of the time, blue ones those for $P$=50\%. 
The red lines (CO type) and green lines (ONe type) show the expected flux for a CO nova at 
the minimum (1.5 kpc) and maximum (12 kpc) of the various published distance estimates  (see Table \ref{17_novae}).
For each binsize, these fluxes were obtained by averaging the appropriate curve in Figure \ref{lc_Hernanz_ONe_CO} from the beginning of the outburst up to a time equal to the binsize.
The purple curves represent the best possible 5$\sigma$ BAT sensitivity, calculated from Equation \ref{sensitivity_eqn} using a typical value of $r_{B}(E)$, if the source were seen continously in the fully coded FOV.

It can be seen that  under ideal conditions  the nova should have been detectable in the higher energy bands if its distance was toward the lower end of the range of estimates.  However the actual upper limits are about an order of magnitude higher than the ideal ones because the nova was in the BAT FOV only for a fraction of the time and then in regions towards the edge having reduced sensitivity. Thus the non-detection is not surprising.

Table \ref{17_novae_final_table} gives, for each of the 24 novae examined, the best constraining timescale ($\Delta t_{best}$), the measured 5$\sigma$ upper limits on the flux, and an estimate of the maximum distance out to which a CO nova should have been detectable ($d_{max}$). The limits are for the 50$-$100 keV band, which was the  best constraining one in each case. When the nova type is unknown, the limiting distance for a ONe nova is indicated into parentheses. The limits and distances are given for two cases - those that can be placed for at least 10\% or for at least 50\% of the possible times of explosion. In no case was the  non-detections surprising, though for several  a detection might have been marginally possible in the most favourable conditions -- if the distance was at the lower end of the range of estimates and the time of explosion were luckily placed with respect to the observations.

\begin{table*}
\caption{\em Summary of the results obtained from BAT data for 24 Classical novae.}
\label{17_novae_final_table}
\begin{center}
\scriptsize{
\begin{tabular}{lccccc|ccc|ccc}
\hline \\
 &  & & & & & \multicolumn{3}{c}{10\% coverage time} & \multicolumn{3}{c}{50\% coverage time} \\
Nova & year & $m_{V}$ & $t_{2}$ & E(B-V) & $d_{MMRD}$ & $\Delta t_{best}$ & Upper limit & $d_{max}$ & $\Delta t_{best}$ & Upper limit  & $d_{max}$ \\
     & & at visual max        &    (days)           &          &    (kpc)       &                       &       (ph cm$^{2}$ s$^{-1}$)                &  (kpc)   &    &  (ph cm$^{2}$ s$^{-1}$)  &  (kpc)   \\
\hline \\
\object{V2361 Cyg} & 2005 &10 &    5.5 $\pm$ 0.5 &      1.2      &          10.80$^{+2.91}_{-2.30}$ & DPH & 5.9 $\times$ 10$^{-3}$ & 3.8       & DPH & 1.5 $\times$ 10$^{-2}$ & 2.3       \\[6pt] 
\object{V382 Nor} &  2005 & 9.7  &  14.5 $\pm$ 2.5 &     1.50-1.51 (*)  &        4.8$^{+1.8}_{-1.2}$ & DPH & 2.1 $\times$ 10$^{-2}$ & 1.8 (0.5) & DPH & 2.6 $\times$ 10$^{-2}$ & 0.1 (0.1) \\[6pt]
\object{V378 Ser} &  2005 & 11.6 &  52  $\pm$ 18 &         0.74     &        18.88$^{+7.71}_{-4.50}$ & DPH & 1.1 $\times$ 10$^{-2}$ & 2.7       & DPH & 4.7 $\times$ 10$^{-2}$ & 0.1       \\[6pt]
\object{V5115 Sgr} &  2005 & 7.75 &  4  $\pm$  2 &        0.53    &          10.22$^{+3.08}_{-2.38}$ & DPH & 6.0 $\times$ 10$^{-3}$ & 3.7 (0.9) & DPH & 2.4 $\times$ 10$^{-2}$ & 0.2 (0.2) \\[6pt]
\object{V1663 Aql} & 2005 & 10.7 &  14 $\pm$  6 &         2       &           3.88$^{+1.83}_{-1.38}$ & DPH & 4.5 $\times$ 10$^{-3}$ & 4.1 (1.0) & 12h & 2.7 $\times$ 10$^{-3}$ & 0.1 (0.2) \\[6pt] 
\object{V5116 Sgr} & 2005 & 7.2  &  7  $\pm$  4 &        0.34-0.57 (*)    &      8.3$^{+5.1}_{-3.1}$ & DPH & 1.2 $\times$ 10$^{-2}$ & 0.1 (0.1) & DPH & 1.2 $\times$ 10$^{-2}$ & 0.1 (0.1) \\[6pt]
\object{V1188 Sco} & 2005 & 8.9 &   12 $\pm$  5 &        1.09-1.49 (*) &         4.4$^{+4.8}_{-1.9}$ & DPH & 5.2 $\times$ 10$^{-3}$ & 3.9 (1.0) & DPH & 5.6 $\times$ 10$^{-2}$ & 0.6 (0.2) \\[6pt]
\object{V1047 Cen} & 2005 & 7.4 &   4.5 $\pm$ 1.5 &      1.28-1.38 (*)  &              2.8 $\pm$ 0.5 & DPH & 1.0 $\times$ 10$^{-2}$ & 2.7 (0.7) & DPH & 2.6 $\times$ 10$^{-1}$ & 0.2 (0.2) \\[6pt]
\object{V476 Sct}  & 2005 & 11.4  &  12 $\pm$ 2 & 2.0                  &      5.69$^{+1.85}_{-1.43}$ & 1h  & 2.4 $\times$ 10$^{-2}$ & 1.8 (0.5) & DPH & 2.4 $\times$ 10$^{-1}$ & 0.1 (0.1) \\[6pt]
\object{V477 Sct}  & 2005 & 10.75 &  7.5 $\pm$ 2.5 &       1.3 &            12.74$^{+4.04}_{-3.16}$ & 1h  & 9.0 $\times$ 10$^{-3}$ & 2.8 (0.7) & 1h  & 5.0 $\times$ 10$^{-2}$ & 0.3 (0.2) \\[6pt]
\object{V2575 Oph} & 2006 & 11 &    31 $\pm$  2 &         1.5 &               5.63$^{+1.69}_{-1.28}$ & DPH & 7.8 $\times$ 10$^{-3}$ & 3.2       & DPH & 4.1 $\times$ 10$^{-2}$ & 1.2       \\[6pt] 
\object{V5117 Sgr} & 2006 & 9.9  &  59 $\pm$  11 &      0.5 $\pm$ 0.15 &     11.91$^{+0.63}_{-0.45}$ & 1h  & 1.3 $\times$ 10$^{-2}$ & 2.5 (0.6) & 1h  & 2.2 $\times$ 10$^{-1}$ & 0.3 (0.2) \\[6pt]
\object{V2362 Cyg} & 2006 & 7.75 &  7  $\pm$  2.5 &       0.59 &              8.90$^{+2.81}_{-2.19}$ & DPH & 1.4 $\times$ 10$^{-2}$ & 2.3       & 1h  & 1.2 $\times$ 10$^{-1}$ & 0.6       \\[6pt]
\object{V2576 Oph} & 2006 & 9.2 &   25.5 $\pm$ 2.5 &      0.62 &              9.58$^{+3.32}_{-2.37}$ & DPH & 8.9 $\times$ 10$^{-3}$ & 3.0       & DPH & 4.6 $\times$ 10$^{-2}$ & 1.3       \\[6pt]
\object{V1065 Cen} & 2007 & 8.7 & 19.5 $\pm$ 1 &  0.77-0.84 (*)       &        7.55$^{+1.65}_{-1.15}$ & 1h  & 7.6 $\times$ 10$^{-3}$ & 3.2 (0.8) & 1h  & 2.7 $\times$ 10$^{-2}$ & 1.6 (0.4) \\[6pt]
\object{V1280 Sco} & 2007 & 4 &   13 $\pm$ 1 &  0.39-0.55 (*)           &        2.1$\pm$0.4  & 1h  & 9.1 $\times$ 10$^{-3}$ & 3.0 (0.8) & DPH & 2.7 $\times$ 10$^{-2}$ & 1.6 (0.4) \\[6pt]
\object{V1281 Sco} & 2007 & 8.8 & 8 $\pm$ 4 &  0.7                   &         12.10$^{+4.28}_{-3.37}$ & 1h  & 9.1 $\times$ 10$^{-3}$ & 2.9 (0.7) & DPH & 3.2 $\times$ 10$^{-2}$ & 1.7 (0.4) \\[6pt]
\object{V2467 Cyg} & 2007 & 7.6 & 8 $\pm$ 2 &  1.6$\pm$0.1                &    1.93$^{+0.26}_{-0.24}$ & 1h  & 9.3 $\times$ 10$^{-3}$ & 2.9 (0.7) & DPH & 5.3 $\times$ 10$^{-2}$ & 1.1 (0.3) \\[6pt]
\object{V2615 Oph}  & 2007 & 8.75 & 36.5 $\pm$ 4.5 &  1.0-1.3               &  3.09$^{+0.21}_{-0.15}$ & 1h  & 2.3 $\times$ 10$^{-2}$ & 1.9       & DPH & 9.3 $\times$ 10$^{-2}$ & 0.9       \\[6pt]
\object{V5558 Sgr}  & 2007 & 6.5 & 6 $\pm$ 1 &  0.8                      &     3.78$^{+1.06}_{-0.83}$ & 1h  & 2.8 $\times$ 10$^{-2}$ & 1.7 (0.5) & DPH & 1.1 $\times$ 10$^{-1}$ & 0.9 (0.2) \\[6pt]
\object{V598 Pup}   & 2007 & ?   & ?         &  ?                       &     ?                     & DPH  & 2.6 $\times$ 10$^{-2}$ & 1.8      & DPH & 1.1 $\times$ 10$^{-1}$ & 0.1       \\[6pt]
\object{V390 Nor}   & 2007 & 9.8 & 49.5 $\pm$ 5.5 &  1.0               &       5.74$^{+1.67}_{-1.26}$ & 1h  & 2.1 $\times$ 10$^{-2}$ & 1.9 (0.5) & 1h  & 6.7 $\times$ 10$^{-2}$ & 1.0 (0.3) \\[6pt]
\object{V458 Vul}   & 2007 & 8.1 & 8.5 $\pm$ 3.5 &  0.6              &     10.01$^{+3.45}_{-2.71}$    & 1h  & 2.8 $\times$ 10$^{-2}$ & 1.7 (0.4) & DPH & 1.1 $\times$ 10$^{-1}$ & 0.4 (0.1) \\[6pt]
\object{V597 Pup}   & 2007 & ? &  ?    &       0.3                     &    ?                           & 1h  & 2.2 $\times$ 10$^{-2}$ & 1.8 (0.5) & DPH & 9.8 $\times$ 10$^{-2}$ & 0.8 (0.2) \\[6pt]
\hline \\
\end{tabular}}
\end{center}
\end{table*}

\subsection{Recurrent and dwarf novae}\label{other_objects}

In parallel with the searches for gamma-ray emission from classical novae, the same technique was applied to certain other objects. This lead to the detection of hard X-ray emission from both the recurrent nova \object{RS Oph} (reported in \citet{Bode}) and the dwarf nova \object{V455 And} \citep{V455_Atel}. The strength and timescale of the high energy outburst from  each of these objects were such that neither was detected on-board or by the routine monitoring on the ground (though more sophisticated analysis ground analysis now being developed might have done so). However a directed search, knowing the exact location and the approximate time of the event, allowed convincing detections.

In each case the strongest emission was below 25 keV with little or no flux at higher energies. This is the inverse of the situation expected for classical novae and the physical processes involved are clearly different. In the case of \object{RS Oph} the emission was due to the shock of the ejecta with the surrounding medium, whereas the emission seen from \object{V455 And} was probably due to an instability in the accretion disc. However these detections demonstrate the potential of retrospective searches such as those described here.

\section{Discussion}

In section \S \ref{mc_results} it was concluded that the BAT should be capable of detecting the gamma-ray emission due to nucleosynthesis in 2$-$5 novae during a 10 yr life.  
This estimate depends on a number of assumptions. The most basic of
these is the predicted gamma-ray flux during nova explosions, which mainly
depends on the expected nucleosynthesis and the hydrodynamics of the
expanding envelope. Indeed the possibility of confirming our current
understanding of nucleosynthesis and dynamical evolution taking place in
nova explosions is one of the incentives of this work.

As explained in \S \ref{Jean_work}, the expected gamma-ray flux from a nova depends critically
on nuclear reactions rates which are still uncertain. Some alternative assumptions could  lead to 
 reduced  expectations and better measurements of certain cross-sections are needed. 
 The WD mass is another important parameter to consider. In this paper we
obtained our predictions assuming that all CO novae and ONe novae 
hosted a 1.15 M$_{\odot}$ or a 1.25 M$_{\odot}$ WD respectively.
In this scenario CO novae are more likely to be detected, but different 
fluxes could be obtained for different WD masses, as shown in
\citet{Hernanz_et_al_2002}. WD masses, as well as the proportion between CO and ONe 
novae in the Galaxy, are very difficult to measure, and only rough estimations
based on scarce information are available.

The results presented in \S \ref{MC_sim} are based on the spatial 
distribution of novae in the Galaxy of model 1 of Table \ref{table_Jean2000}.
It is noteworthy that the conclusions  do not strongly depend on the model chosen; those obtained with models 2 and 3 are very similar. This is because BAT could only detect novae within 
$\sim$5 kpc, where the spatial distributions are similar (see Figure \ref{distrb_distanze_novae_simul_3modelli.ps}).

The other factor that can strongly affect our results is the rate of novae in the Galaxy, or more particularly the rate of discovered novae. It is certain that some relatively nearby novae are never observed, because they were too close to the sun, though chance of cloud coverage, or (particularly in the southern hemisphere)  simply because no one was looking. The estimated discovery probability of 50\% for novae  $m_V <$ 9  given in \S \ref{mc_results} is based on 20 yr of data. The rate of discovery and the methods used are evolving quickly and it is difficult to extract statistics on the distribution and characteristics of those currently being found. Almost certainly the chance that a nova potentially detectable with BAT will be reported is already higher.

 The fact that no novae were detected in the first  3 yr of BAT operations (\S  \ref{results})  is consistent with the expectation  of 2$-$5 nova in 10 years estimated in \S \ref{mc_results}.  None of the 24 novae occurring during that time was sufficiently close and sufficiently well covered that a detection was likely.
 The fact that the technique was able to detect similar outbursts from \object{RS Oph} and \object{V455 And} demonstrates its efficacy even though the emission had a different origin.
 The search is being continued and the analysis presented here gives high hopes of eventual success.

An intriguing possibility is that a sufficiently nearby nova might be detectable without depending on the later detection of an optical nova. If it were found by the on-board software, rather than later on the ground, it might even trigger a spacecraft slew to bring the position within the field of view of the XRT and UVOT instruments. Because the gamma rays are hardly affected by interstellar dust, it could even
lead to the discovery of a nova that would otherwise be unobserved (IR observations would allow the confirmation of the nature of the event). The probability of a nova so close is, of course relatively low, but the information that could be obtained would be of enormous value.

\begin{acknowledgements}

The authors wish to thank members of the \Swift\  BAT team for useful discussions and advice.  We acknowledge with thanks the variable star observations from the AAVSO International Database contributed by observers worldwide and used in this research. This work was supported by the Centre National d'Etudes Spatiales (CNES). It is based on observations with BAT embarked on SWIFT. Financial support from MEC ESP2007-61593, AGAUR 2005-SGR00378 and FEDER funds is also acknowledged. 

\end{acknowledgements}

\bibliographystyle{aa}
\bibliography{paper_novae_26Jan08}

\end{document}